\documentclass[apj]{emulateapj}

\slugcomment{To appear in the {\it Astrophysical Journal}}
\shorttitle{CMEs}
\shortauthors{Drake et al.}

\begin{document}

\title{Implications of mass and energy loss due to coronal mass ejections on magnetically-active stars}

\author{Jeremy
  J.~Drake\altaffilmark{1}, Ofer Cohen\altaffilmark{1},
  Seiji Yashiro\altaffilmark{2,3}, Nat Gopalswamy\altaffilmark{3}}
\affil{$^1$Harvard-Smithsonian Center for Astrophysics,
    60 Garden Street, Cambridge, MA 02138}
\affil{$^2$Interferometrics Inc., Herndon, VA 20171}
\affil{$^3$ NASA Goddard Space Flight Center, Greenbelt, MD 20771}

    
\email{jdrake@cfa.harvard.edu}

\begin{abstract} 
Analysis of a database of solar coronal mass ejections (CMEs) and associated flares over the period 1996-2007 finds well-behaved power law relationships between the 1--8~\AA\ flare X-ray fluence and CME mass and kinetic energy.  We extrapolate these relationships to lower and higher flare energies to estimate the mass and energy loss due to CMEs from stellar coronae, assuming that the observed X-ray emission of the latter is dominated by flares with a frequency as a function of energy $dn/dE=kE^{-\alpha}$.   For solar-like stars at saturated levels of X-ray activity, the implied losses depend fairly weakly on the assumed value of $\alpha$ and are very large:  $\dot{M}\sim 5\times 10^{-10}M_\odot$~yr$^{-1}$ and $\dot{E}\sim 0.1L_\odot$.  In order to avoid such large energy requirements, either the relationships between CME mass and speed and flare energy must flatten for X-ray fluence $\ga 10^{31}$~erg, or the flare-CME association must drop significantly below 1 for more energetic events.   If active coronae are dominated by flares, then the total coronal energy budget is likely to be up to an order of magnitude larger than the canonical $10^{-3}L_{bol}$ X-ray saturation threshold.  This raises the question of what is the maximum energy a magnetic dynamo can extract from a star?  For an energy budget of 1\%\ of $L_{bol}$, the CME mass loss rate is about $5\times 10^{-11}M_\odot$~yr$^{-1}$.  
\end{abstract}



\section{Introduction}
\label{s:intro}

The rate of mass loss from unevolved late-type stars is notoriously difficult to constrain.  In case of the Sun, the wind can be directly observed and sampled by spacecraft, and amounts to a mass loss rate of about $2\times 10^{-14} M_\odot$~yr$^{-1}$. 
Such a weak flow of ionized gas from other stars cannot be detected directly using instrumentation available today, and there are presently no {\it direct} detections or measurements of winds from solar-like stars.   Mass loss rate upper limits based on radio observations are in the range of several $10^{-11}M_\odot$~yr$^{-1}$ \citep[e.g.][]{Gaidos.etal:00}.  \citet{Wood.etal:02} devised a method of indirect assessment based on H~Ly$\alpha$ absorption due to interstellar H~{\sc I} that is heated in the interaction region between the wind and the local interstellar medium. They estimated rates in the range $10^{-15}$--$10^{-12}M_\odot$~yr$^{-1}$, with evidence for higher mass loss rates for more X-ray luminous stars.  

Theoretical progress in predicting solar-like winds has been hampered by a persistent lack of understanding of the basic mechanisms responsible for producing them.  Inspired by the recent success of turbulence-driven coronal heating and solar wind acceleration theory, \citet{Cranmer.Saar:11} developed a wind model based on the energy flux of magnetohydrodynamic turbulence from the subsurface convection zone.  For a solar-like star, they predict mass loss rates that decline steadily from a few $10^{-12}M_\odot$~yr$^{-1}$ at the zero-age main-sequence and rotation periods of 1~day to $\sim 10^{-15}M_\odot$~yr$^{-1}$ at rotation periods of 60 days or so, in reasonable agreement with the estimates of \citet[][see also \citealt{Wood.etal:05}]{Wood.etal:02}.

Further progress in understanding mass loss of main-sequence late-type stars is strongly motivated by the effect winds have on stellar rotation evolution \citep[e.g.][]{Weber.Davis:67,Stauffer.Hartmann:86,Kawaler:88,Matt.Pudritz:08,Reiners.Mohanty:12} 
and consequently on stellar magnetic activity \citep[e.g.][]{Pallavicini.etal:81,Wright.etal:11},  and on interplanetary medium environments \citep[e.g.][]{Preusse.etal:05,Lammer.etal:07}.  

One aspect of mass loss that remains to be thoroughly investigated on other late-type stars is that due to coronal mass ejections (CMEs).  On the Sun, CMEs are observed to eject from $10^{13}$ to $10^{17}$g of magnetized plasma into the interplanetary medium \citep[e.g][]{Yashiro.Gopalswamy:09,Vourlidas.etal:10}.  The integrated mass loss from CMEs can amount to several percent of the steady wind rate \citep[e.g][]{Vourlidas.etal:10}.  At first sight this suggests CMEs are going to be of little importance in the stellar context.  However, on the Sun CMEs are associated with flares, and magnetically active stars are widely interpreted to be dominated by flares \citep[e.g.][]{Guedel:97,Drake.etal:00}.   Since the most magnetically active solar-like stars can attain X-ray luminosities more than 1000 times that of the Sun, there is scope for vigorous CME activity, especially in the context of recent giant flare detections on solar-type stars based on optical {\it Kepler} photometry by \citet{Maehara.etal:12}.   The importance of CMEs on active stars has also been raised in the context of erosion of the atmospheres of ``Hot Jupiters'' \citep{Khodachenko.etal:07} and the habitability of planets around M dwarfs \citep{Khodachenko.etal:07b}, while 
\citet{Aarnio.etal:12} suggest that CMEs associated with flares could be an important contribution to angular momentum loss on pre-main sequence stars.

Here, we examine the consequences for mass loss and energy loss of flare-dominated coronae based on extrapolation of the observed behavior of a large sample of CMEs compiled by \citet{Yashiro.Gopalswamy:09}.  The implications are quite striking and provide an indirect means to begin to assess how CMEs might behave on stars much more magnetically-active than the Sun.

\section{Mass and Energy of Solar Coronal Mass Ejections}
\label{s:cmes}

As a guide to the CME behaviour of active stars we look to the Sun.  \citet{Yashiro.Gopalswamy:09} studied the statistical relationships between solar flares and CMEs observed over the period 1996-2007 (see also \citealt{Vourlidas.etal:10} for a description of some of the observational aspects of CME measurements).  They compiled a database of soft X-ray flares observed by the Geostationary Operational Environmental Satellite (GOES) that were associated with CMEs observed by the Large Angle and Spectrometric Coronagraph (LASCO) on board the Solar and Heliospheric Observatory (SoHO) mission.  The CME-flare association fraction was observed to increase with flare peak X-ray flux, fluence, and duration, as had been noted in earlier studies \citep[e.g.][]{Andrews:03}, and a good correlation was found between the flare fluence and the CME kinetic energy. 

\begin{figure}
\epsscale{1.2}
\plotone{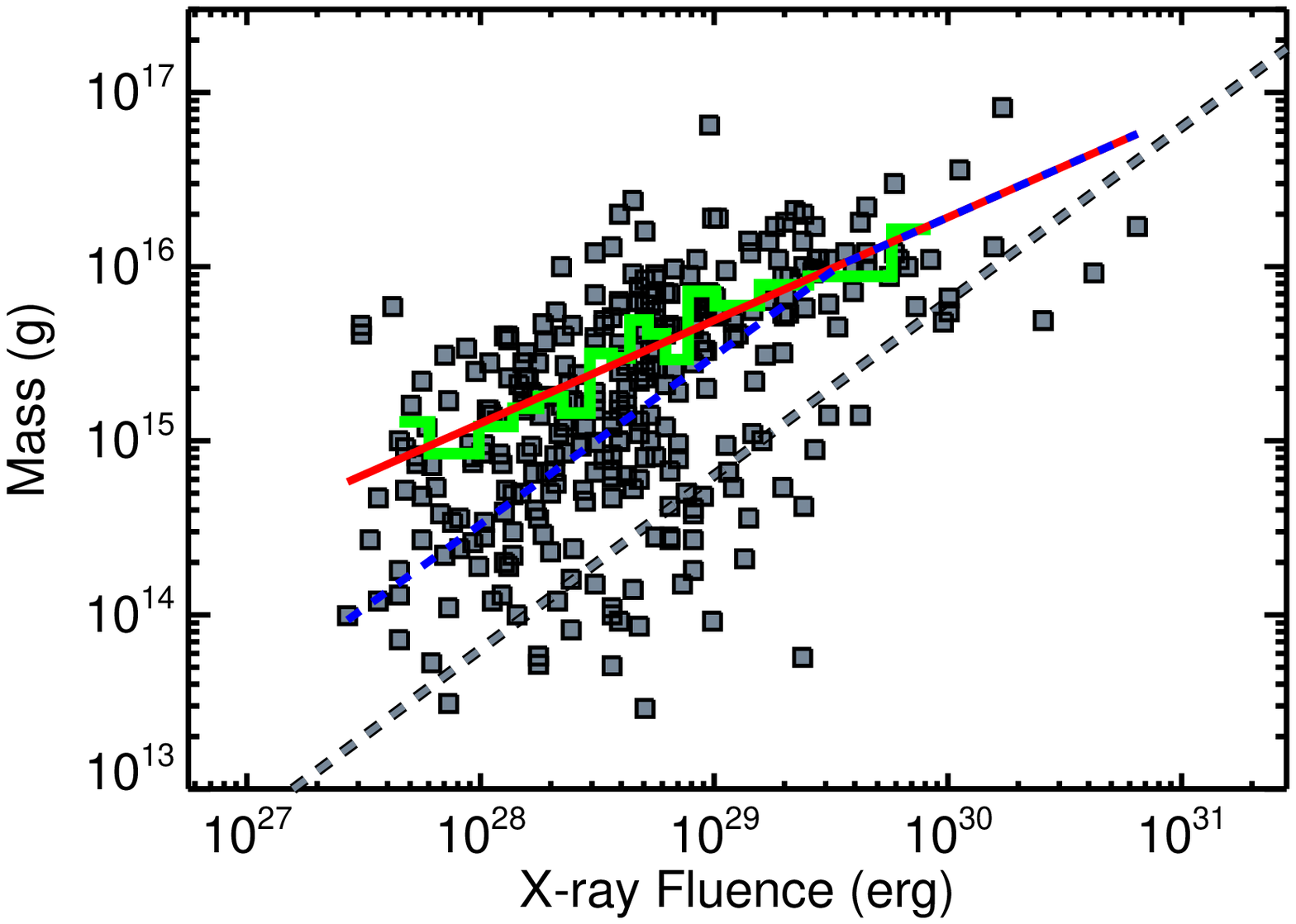}
\plotone{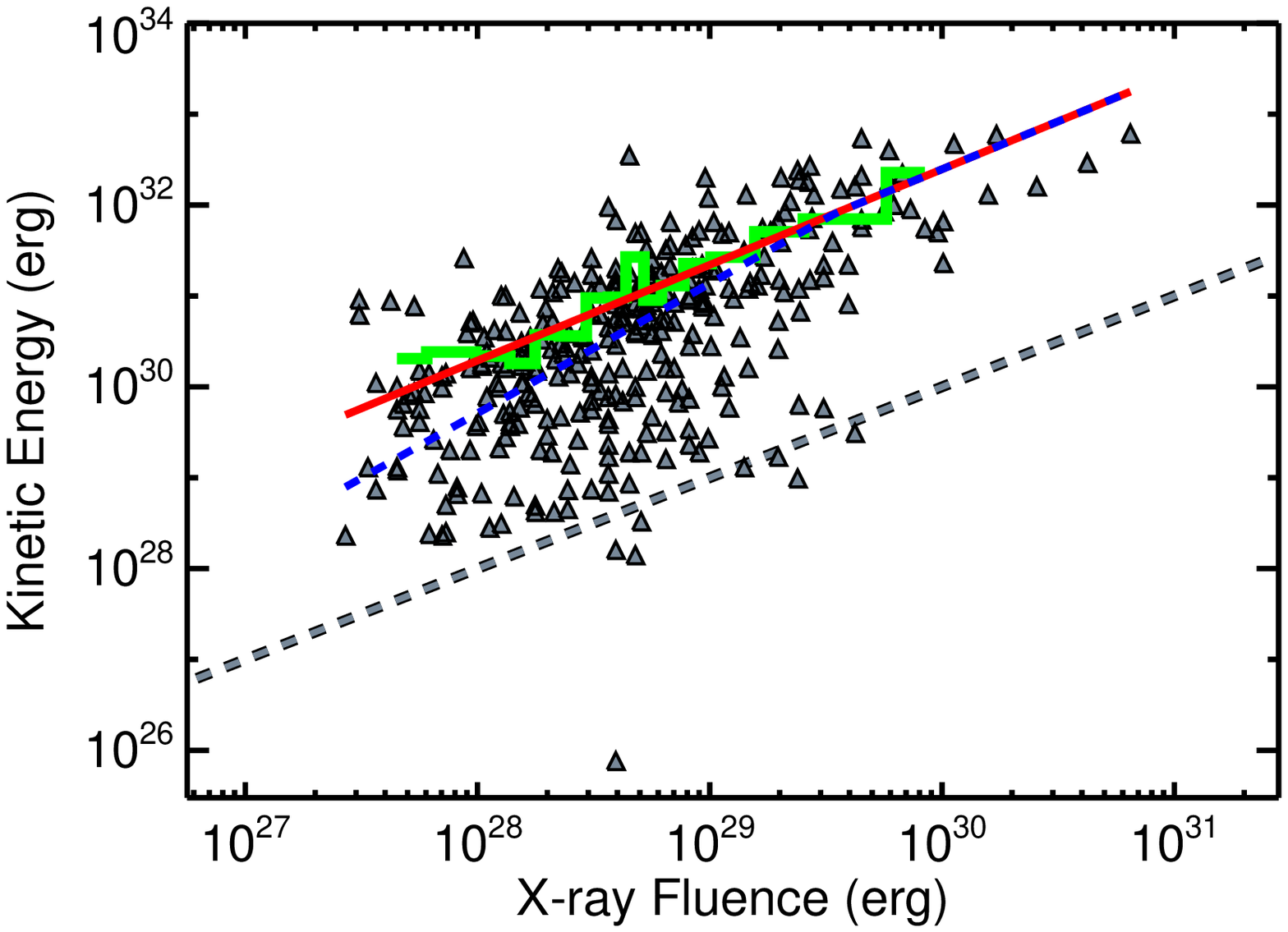}
\caption{CME mass (top) and kinetic energy (bottom) vs X-ray fluence of the associated flare from the \citet{Yashiro.Gopalswamy:09} sample.  The green histograms are the means over twenty data points and the red lines are linear fits to these means.  
The dashed blue lines are the linear fits multiplied by the CME-flare association rate given by Eqn.~\ref{e:ass_frac}.  In the upper panel, the dashed grey line follows a constant ratio of mass loss to GOES X-ray energy loss, expressed as rates, $\dot{M}=10^{-10}(L_X/10^{30}) M_\odot$~yr$^{-1}$.
In the lower panel, the dashed grey line represents equivalence of the kinetic and X-ray energies.  The red line in this panel corresponds very closely to a factor of two hundred times the X-ray fluence.  
\label{f:cme_lx}}
\end{figure}

The distribution of CME ejected masses as a function of the associated flare GOES 1--8~\AA\ X-ray fluence from the \citet{Yashiro.Gopalswamy:09} sample is illustrated in Figure~\ref{f:cme_lx}.   While the data show a very large scatter in CME mass at a given flare energy, we find that the mean of these data over small X-ray fluence bins is well-behaved and adheres quite closely to a power law.  Such a power law relation between ejected mass and the peak X-ray flux of the associated flare has also been pointed out by \citet{Aarnio.etal:11b}.   A linear fit to the logarithm of the variance-weighted means of 20 point bins yields the following power law relationship (in cgs units) between ejected mass and flare fluence,
$$
m_c(E)=  \mu E^\beta \, ;
$$
\begin{equation}
\mu=10^{-1.5\mp 0.5}, \,\,\, \beta=0.59\pm 0.02. 
\label{e:cme_mass}
\end{equation}
The constant of proportionality and power law index uncertainties are strongly anti-correlated.  They were determined using a Monte Carlo multiple imputation bootstrap \citep{Rubin:96} in which the distribution of fit parameters was estimated from repeated re-fitting of a randomly-drawn 2/3 of the data augmented to the full sample size by random draws from this sub-sample. 

\citet{Yashiro.Gopalswamy:09} found a similar distribution of CME kinetic energy as a function of associated flare X-ray fluence.   The data are shown in Figure~\ref{f:cme_lx}, together with the histogram of the mean of every 20 points and a power law fit to this.   From this fit we find the mean CME kinetic energy to vary with flare X-ray fluence as (again in cgs units)
$$
E_{ke}= \eta E^\gamma \, ;
$$
\begin{equation}
\eta=10^{0.81\mp 0.85}, \,\,\, \gamma=1.05\pm 0.03. 
\label{e:cme_ke}
\end{equation}
Again, the uncertainties  were determined using a Monte Carlo multiple imputation bootstrap and are essentially anti-correlated.
Also shown in Figure~\ref{f:cme_lx} is the locus of equivalence between X-ray and kinetic energies.   The latter lies about a factor of 200 above the former, indicating that, for a given flare, the energy release will be totally dominated by the energy of the associated mass ejection.  We return to the consequences of this in \S\ref{s:discuss}. 

We will find it useful below to also express the CME-flare association fraction as a power law.   We find the \citet{Yashiro.Gopalswamy:09} association fraction as a function of X-ray fluence, $f(E)$, can be well-represented by 
$$
f(E)= 1 \,\,\, {\rm for}\,\,\, E > 3.5\times 10^{29} \, {\rm erg}
$$
$$
f(E)= \zeta E^\delta \,\,\, {\rm for}\,\,\,E \leq 3.5\times 10^{29} \, {\rm erg} \, ; 
$$
\begin{equation}
\zeta=7.9\times10^{-12}, \,\,\, \delta=0.37. 
\label{e:ass_frac}
\end{equation}

\section{Estimating total stellar CME-associated energy and mass loss}
\label{s:method}

Flare occurrence in the coronae of the Sun and stars has been shown by a number of studies to follow a power law distribution in frequency as a function of flare energy of the form
\begin{equation}
\frac{dn}{dE}=k E^{-\alpha},
\label{e:hudson}
\end{equation}
where $k$ is a normalization constant \citep[e.g.][]{Drake:71,Datlowe.etal:74,Lin.etal:84,Hudson:91,Bai:93,
Porter.etal:95,Krucker.Benz:98,Audard.etal:00,Kashyap.etal:02,Gudel.etal:03,Hannah.etal:11}.
Analyses of stellar EUV and X-ray light curves furthermore suggest active stellar coronae are dominated by a superposition of flares, and tend to find values of the frequency vs.\ energy power law index in the range $\alpha=2$--2.5 for all stellar types, including dwarfs with spectral type G-M \citep[][see also the earlier work of \citealt{Collura.etal:88}]{Audard.etal:00,Kashyap.etal:02,Gudel.etal:03,Telleschi.etal:05}, and T~Tauri stars \citep{Caramazza.etal:07,Stelzer.etal:07}.   Similar indices were observed for optical flares detected on solar-type stars by \citet{Maehara.etal:12} from {\it Kepler} photometry.  

The total flare power  is given by the integral over minimum and maximum flare energies 
\begin{equation}
P=\int_{E_{min}}^{E_{max}} E k E^{-\alpha}\, dE = \frac{k}{2-\alpha}\left[E_{max}^{2-\alpha} - E_{min}^{2-\alpha}\right].
\label{e:power}
\end{equation}
Taking the X-ray luminosity, $L_X$, as the observable proxy for the flare power, the constant $k$ is given by 
\begin{equation}
k=\frac{L_X (2-\alpha)}{\left(E_{max}^{2-\alpha} - E_{min}^{2-\alpha}\right)}.
\label{e:constant}
\end{equation}

For the case in which CME mass loss is a function of the associated flare X-ray fluence, the total mass loss rate from CMEs is 
\begin{equation}
\dot{M}_c=\int_{E_{min}}^{E_{max}} m_c(E) f(E) \frac{dn}{dE} \, dE.
\end{equation}
When combined with Eqns.~\ref{e:cme_mass}, \ref{e:ass_frac} and \ref{e:constant}, the resulting rate is 
\begin{equation}
\dot{M}_c=\mu \zeta L_X \left(\frac{2-\alpha}{1+\beta+\delta-\alpha}\right) \left[\frac{E_{max}^{1+\beta+\delta-\alpha}-E_{min}^{1+\beta+\delta-\alpha}}{E_{max}^{2-\alpha}-E_{min}^{2-\alpha}}\right] ,
\label{e:mcme}
\end{equation}
which we can now evaluate for suitable choices of the minimum and maximum flare energies.   Similarly, the total CME-associated kinetic energy loss rate is 
\begin{equation}
\dot{E}_{ke}=\eta \zeta L_X \left(\frac{2-\alpha}{1+\gamma+\delta-\alpha}\right) \left[\frac{E_{max}^{1+\gamma+\delta-\alpha}-E_{min}^{1+\gamma+\delta-\alpha}}{E_{max}^{2-\alpha}-E_{min}^{2-\alpha}}\right].
\label{e:kecme}
\end{equation}

\begin{figure}
\epsscale{1.2}
\plotone{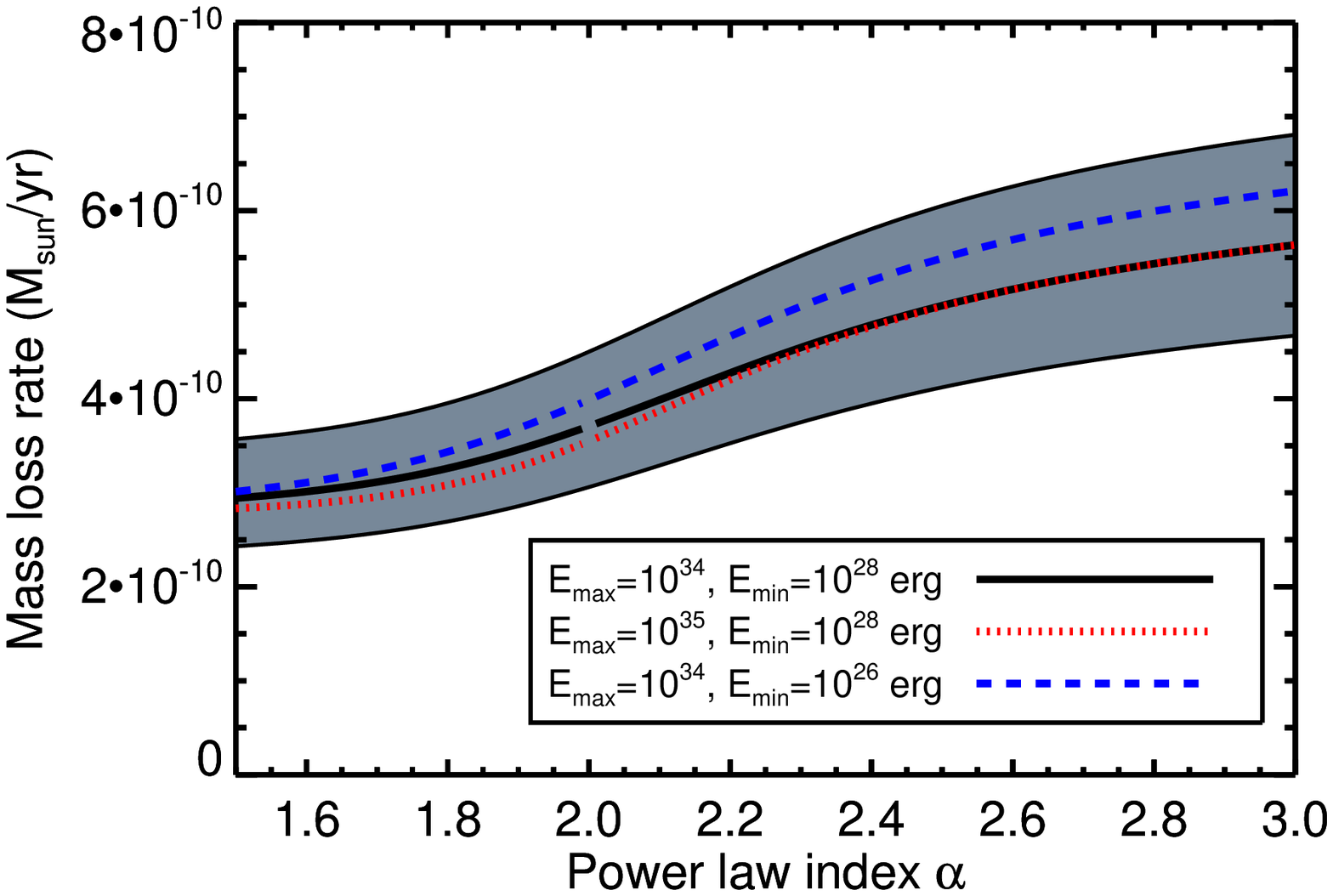}
\plotone{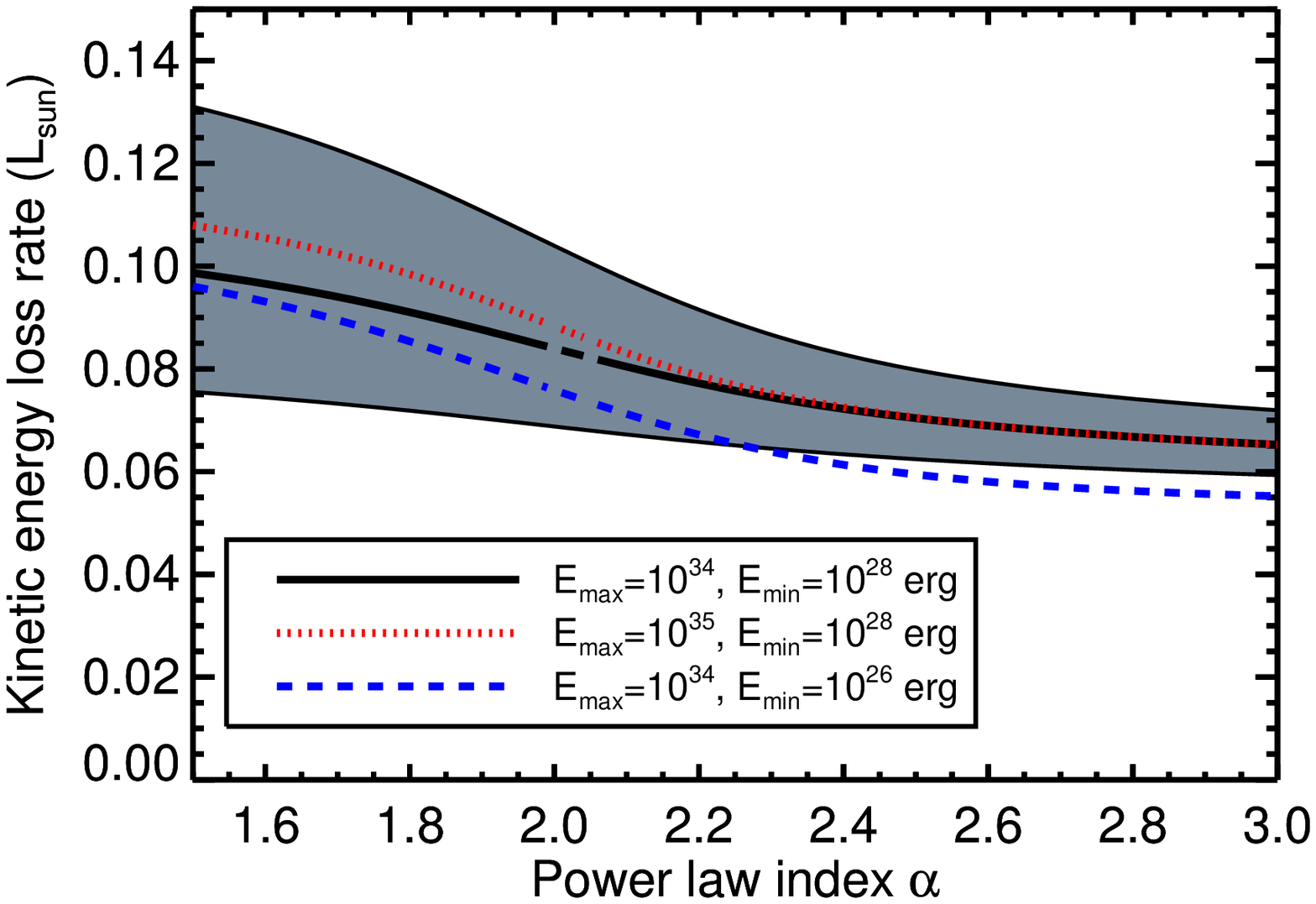}
\caption{CME mass (top) and kinetic energy (bottom) loss rates vs.\  power law index $\alpha$ for a 1--8~\AA\ X-ray luminosity of $L_X=10^{30}$~erg~s$^{-1}$, according to Eqns.~\ref{e:mcme} and \ref{e:kecme}, respectively.  Note that there are singularities corresponding to power law indices $\alpha=1+\beta+\delta$ and $\alpha=2$.  The grey shaded areas represent the uncertainties in the loss rates corresponding to the uncertainties in the power law fits in Eqns.~\ref{e:cme_mass} and \ref{e:cme_ke}.
\label{f:alpha_loss}}
\end{figure}

Mass loss and kinetic energy loss rates are illustrated as a function of the power law index $\alpha$ in Figure~\ref{f:alpha_loss} for different values of $E_{min}$ and $E_{max}$.    Here, we have normalised to a coronal luminosity of $L_X=10^{30}$~erg~s$^{-1}$ in the GOES 1--8~\AA\ bandpass, which, for a fairly typical coronal temperature for the most active solar-like stars of $2\times10^7$~K,  corresponds to $L_X\sim 3\times 10^{30}$~erg~s$^{-1}$ in the 0.2--2.5~keV and 0.5--10~keV bandpasses.   This X-ray luminosity is that of a coronally-saturated solar-like star with a ratio of X-ray to bolometric luminosity of $L_X/L_{bol}\sim 10^{-3}$, such as 47~Cas~B or EK~Dra 
\citep[see, e.g.,][]{Telleschi.etal:05}.

For fiducial limits we have adopted $E_{max}=10^{34}$~erg and $E_{min}=10^{-6}E_{max}$.  The former corresponds to a reasonably large but fairly common flare on an active solar-type star. 
Figure~\ref{f:alpha_loss} demonstrates that the particular choice of these integration limits is not important---changing the lower limit by a factor of 100, for example, barely affects the derived mass loss rate and changes the kinetic energy by an amount comparable to the uncertainty resulting from the power law fit in Eqn.~\ref{e:cme_ke}.   The general conclusion from Figure~\ref{f:alpha_loss} is that, for values of $\alpha\sim 2$--2.5, the CME mass loss rate for a saturated solar-type star is $\dot{M}\sim 5\times 10^{-10}M_\odot$~yr$^{-1}$.  The corresponding CME kinetic energy requirement approaches $\dot{E}_{ke}\sim 0.1L_\odot$.   In the context of current ideas concerning mass loss and efficiency of magnetic energy dissipation on active late-type stars these values are extremely high and their implications are discussed in \S\ref{s:mdot} and \S\ref{s:ke}.

The implied CME mass and energy loss rates in stars with magnetic activity significantly below the saturation threshold depend much more heavily on the X-ray bandpasses and the luminosity that enters the normalisation factor in Eqns.~\ref{e:mcme} and \ref{e:kecme}.   We use a relation between coronal temperature and 0.1--10~keV X-ray luminosity for solar-like stars based on those of \citet{Guedel.etal:97} and \citet{Telleschi.etal:05}, 
\begin{equation}
L_X=6\times 10^{25} \widetilde{T}^{4.5} \,\, {\rm erg\; s^{-1}},
\label{e:lx_t}
\end{equation}
where the constant and power law index have been tailored slightly so as to represent the isothermal plasma temperature, $\widetilde{T}$, that reproduces the observed hardness ratios $\frac{L(1.0-10keV)}{L(0.2-1.0keV)}$ 
derived by \citet{Telleschi.etal:05}.   The hard band adopted by those authors is similar to the GOES band (1.54--12.4~keV) and the relation in Eqn.~\ref{e:lx_t} provides a reasonably accurate means for scaling the broad-band X-ray luminosity to this harder bandpass.   Model hardness ratios and bandpass conversion factors as a function of temperature were derived using the APEC radiative loss model, as implemented in PIMMS\footnote{http://cxc.harvard.edu/toolkit/pimms.jsp}.   The resulting scaling factor depends approximately linearly on $\log{L_X}$, varying from $\sim 0.3$ at $L_X=3\times 10^{30}$~erg~s$^{-1}$ to $10^{-3}$ at $L_X=10^{27}$~erg~s$^{-1}$.

\begin{figure}
\epsscale{1.2}
\plotone{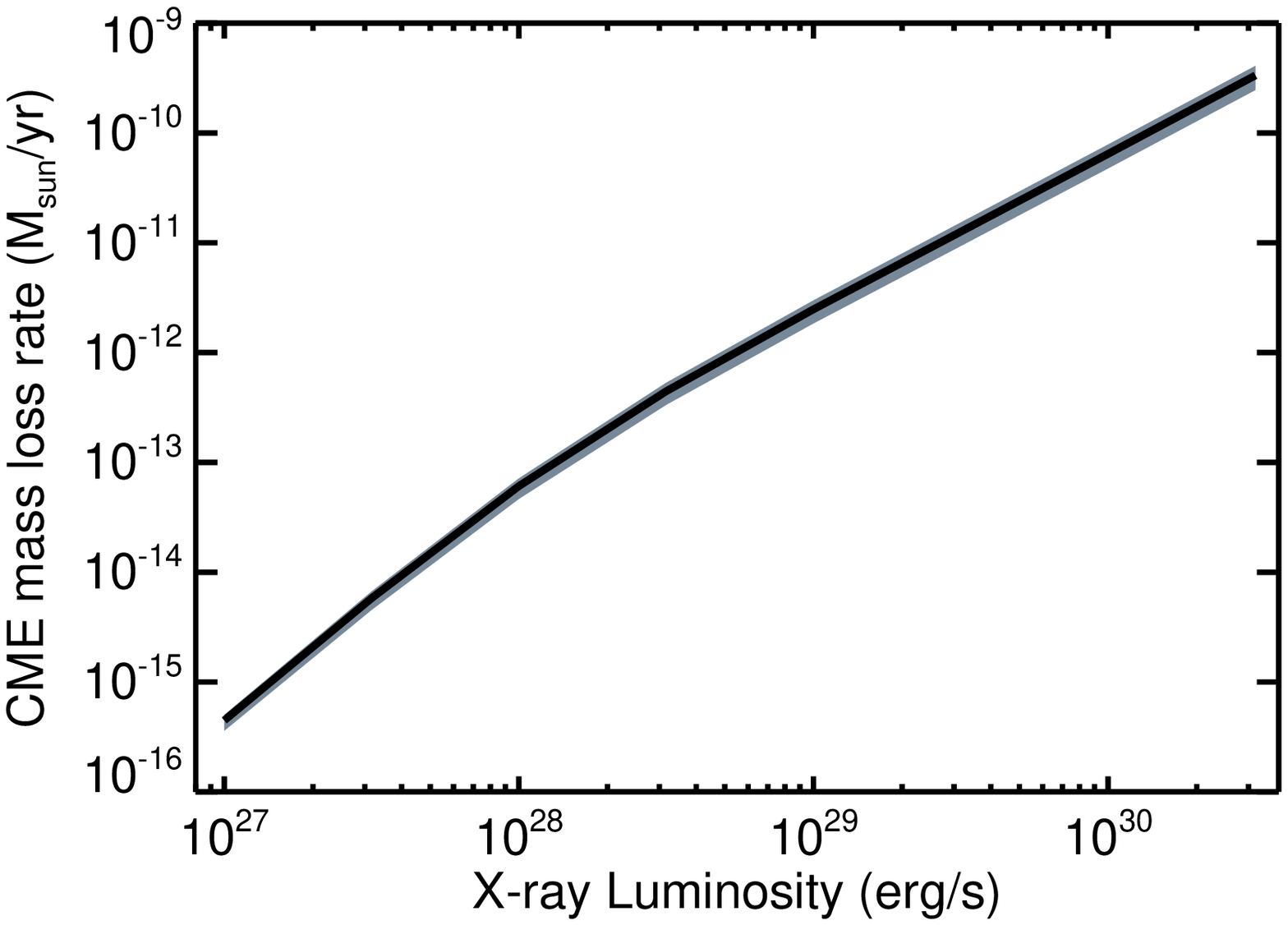}
\plotone{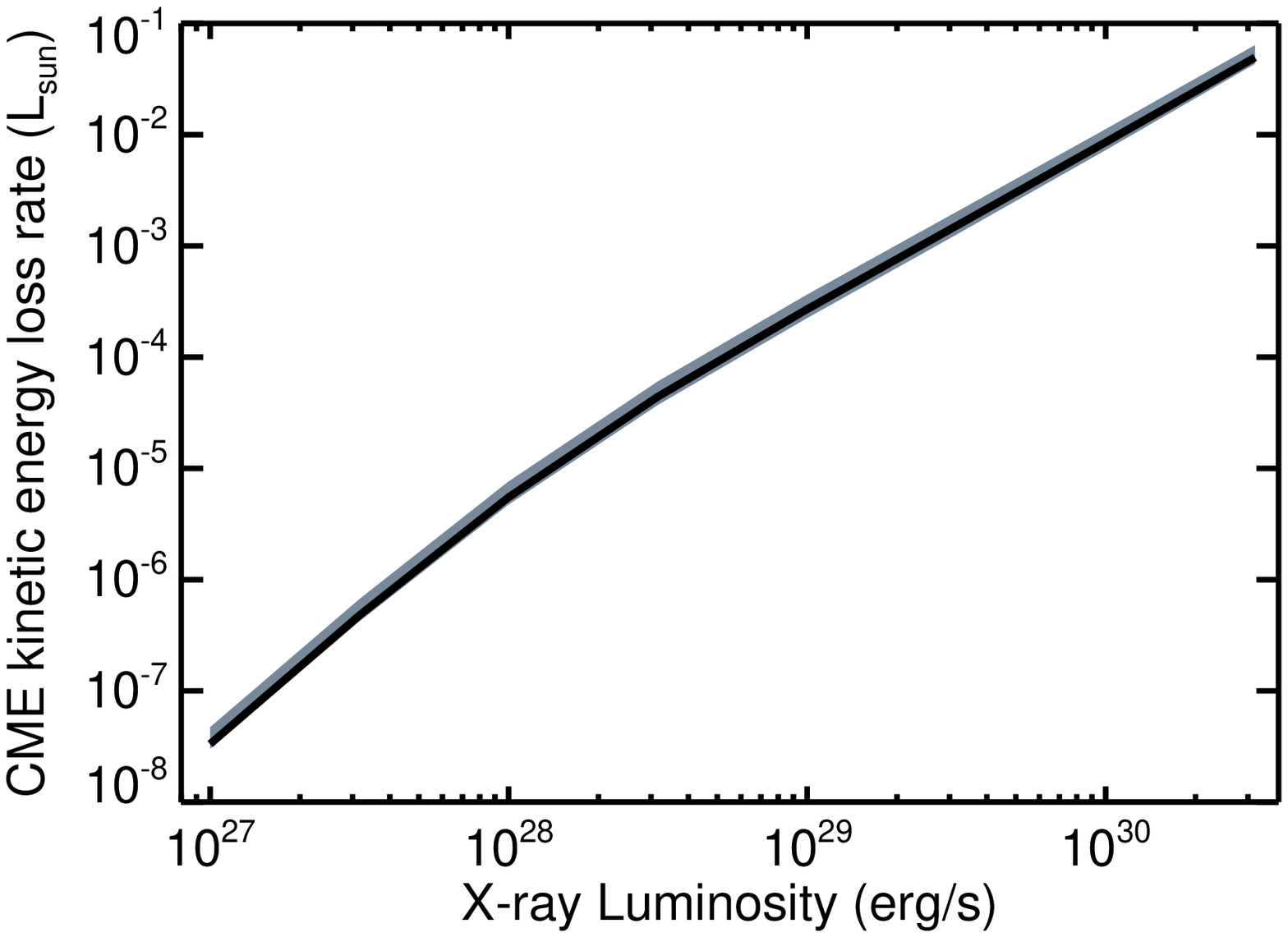}
\caption{CME mass (top) and kinetic energy (bottom) loss rates vs.\  broad-band X-ray luminosity. The solid curve represents the power law index $\alpha=2.25$.  The grey shaded areas represent the range in the loss rates corresponding to the power law index range $1.5 \leq \alpha \leq 3.0$.
\label{f:lx_loss}}
\end{figure}

Using the $L_X$ to GOES bandpass scaling factor, we can obtain the CME mass and energy loss rates as a function of broad-band X-ray luminosity.  These are illustrated in Figure~\ref{f:lx_loss}.  Here, the minimum and maximum flare energies were assumed to be $E_{min}=10^{-6}E_{max}$ and $E_{max}=10^4L_X$, though again the results depend only weakly on the exact limits of integration.  The CME mass loss rate as a function of X-ray luminosity can be approximated by a simple power law at higher energies and a polynomial over a larger energy range:
$$
\log{\dot{M}_c}= -54.6 + 1.48\log{L_X} \,;\,\,\, L_X \ge 10^{28}\,\,{\rm erg~s^{-1}} 
$$
\begin{equation}
\log{\dot{M}_c}= -1339 + 131\log{L_X} -4.37\log^2{L_X}+0.049\log^3{L_X} 
\label{e:cmempl}
\end{equation}
where the latter third order relation should be valid for $L_X$ down to $10^{26}$~erg~s$^{-1}$. 

\section{Discussion}
\label{s:discuss}

The reader familiar with the literature touched upon in \S\ref{s:intro} on winds from late-type main-sequence stars might view the mass loss rates derived in \S\ref{s:method} with incredulity.   The CME kinetic energy loss rate approaching 1/10th of the stellar luminosity also seems implausibly high when compared with radiative losses through X-rays for saturated stars.  For a range of spectral types this saturation level is consistently close to $10^{-3}L_{bol}$, with a scatter of a factor of 2--3 \citep[e.g.][]{Wright.etal:11}.   
The derivations themselves in \S\ref{s:method} are, however,  completely straightforward and the origin of the numbers is easy to understand simply though inspection of the solar CME data.  
At face value, uncertainties in the analysis resulting from the power law fitting are also fairly small, resulting in mass and kinetic energy loss uncertainties of factors of a few.  However, it should be kept in mind that there are relatively few CMEs at the higher energy and mass end of the observed distribution and systematic uncertainties resulting from this likely dominate.

\subsection{Mass loss}
\label{s:mdot}

Also shown in Figure~\ref{f:cme_lx} is the vector corresponding to a constant ratio of mass loss to GOES X-ray energy loss converted to loss rates, $\dot{M}=10^{-10}(L_X/10^{30}) M_\odot$~yr$^{-1}$, and the mean CME ejected mass vs.\ flare X-ray fluence weighted by the CME-flare association fraction.  For flare energies $\leq 3.5 \times 10^{29}$~erg, the latter lies remarkably parallel to the former (probably by coincidence, but if not it raises an interesting issue), and offset by a factor of 5 or so.   One can see that this translates directly to the derived mass loss rate for a flare-dominated corona of  a few $10^{-10}M_\odot$~yr$^{-1}$, as found in \S\ref{s:method} and shown in Figure~\ref{f:alpha_loss}, with little dependence on the power law index $\alpha$.   

The value of $\alpha$ essentially controls the weighting between flares of lower or higher energy.  This is evident from Eqn.~\ref{e:power}: for $\alpha <2$, the X-ray luminosity tends to be dominated by the larger flares, whereas for $\alpha >2$ smaller flares contribute the largest fraction to the observed emission.    Since the flare association-weighted mean CME mass follows very closely a constant $E$ vs.\ mass, the total CME mass loss depends only weakly on  $\alpha$.  Above the energy at which  the CME-flare association is unity ($3.5 \times 10^{29}$~erg), the slope of the $E$-mass  relation is more shallow and we expect the total derived mass loss to be  lower  for $\alpha < 2$, as is borne out in Figure~\ref{f:alpha_loss}.

It is of interest to see how the CME scaling might apply to solar levels of activity.  From Figure~\ref{f:lx_loss}, we find that, for a fairly active solar X-ray luminosity of $L_X\sim 10^{27}$~erg~s$^{-1}$, the mass loss rate is $\dot{M}c\sim 4\times 10^{-16} M_\odot$~yr$^{-1}$---a few percent of the total solar mass loss rate, in broad agreement with the average CME mass flux assessed by \citet{Vourlidas.etal:10} between 1999 and 2003.

The conclusion from the mean $E$ vs.\ mass relation for solar flares and associated CMEs is that, if active stars are dominated by solar-like flares and we can extrapolate the relation in Figure~\ref{f:cme_lx} to higher flare energies, CME mass loss rates are very large---up to four orders of magnitude greater than the present-day solar wind and scaling with 1--8~\AA\ band X-ray luminosity roughly according to $\dot{M}\sim 5\times 10^{-10}(L_X/10^{30}) M_\odot$~yr$^{-1}$.   \citet{Aarnio.etal:12} find similar numbers for T~Tauri stars ($7\times 10^{-11}$--$2\times 10^{-9} M_\odot$~yr$^{-1}$) using a similar method to that employed here.

Can such high mass loss rates possibly be correct?  For the most active stars, the CME energy requirements clearly pose a problem; we return to this in \S\ref{s:ke}. 
In \S\ref{s:intro} we cited evidence that upper limits to wind-driven mass loss in active solar-type stars based on attempts to detect free-free radio emission were of the order of a few $10^{-11}M_\odot$~yr$^{-1}$ \citet{Gaidos.etal:00}.   This analysis assumed a spherically-symmetric wind, but a superposition of many CMEs should produce a qualitatively similar, though probably more clumpy and turbulent, outflow (see also \S\ref{s:fys}).  The most active star in the sample was $\pi^1$~UMa, with $\dot{M} \leq 5\times 10^{-11}M_\odot$~yr$^{-1}$.  This star has a broad-band X-ray luminosity of 
$L_X\sim 10^{29}$~erg~s$^{-1}$ \citep{Drake.etal:94,Telleschi.etal:05}---an order of magnitude or so below the saturation level.     
Based on our $\dot{M}_c$ vs.\ $L_X$ relation in Figure~\ref{f:lx_loss}, this implies $\dot{M}\sim 3\times 10^{-12} M_\odot$~yr$^{-1}$---well within the \citep{Gaidos.etal:00} upper limit.   This is also consistent with the maximum allowed mass loss rate of $\leq 10^{-11}M_\odot$ for a $10^7$~K ``wind'' estimated by \citet{Lim.White:96} based on the requirement of radio transparency consistent with radio detections of active stars.  

\citet{Wood.etal:02} and \citet{Wood.etal:05} estimated mass loss rates for a handful of stars of different activity level using astrospheric Ly$\alpha$ absorption.  While their analysis assumed a spherically-symmetric outflow with a wind speed of 400~km~s$^{-1}$, we would again expect a similar observational signature from a superposition of CMEs.  The Wood et al.\ measurements scale with wind ram pressure, so an outflow comprised of generally faster CMEs would imply a proportionately lower mass loss rate.   The mass-weighted mean CME speed in the \citet{Yashiro.Gopalswamy:09} sample is 1015~km~s$^{-1}$, implying a lower mass loss by only a factor of $\sim 2$.
For a small sample of G and K dwarfs \citet{Wood.etal:05} found $\dot{M}\propto F_X^{1.34\pm {0.18}}$, where $F_X$ is the surface X-ray flux.   This is similar to the CME mass loss power-law relation we found in Eqn.~\ref{e:cmempl}.
While they caution against  extrapolating the relation to higher activity levels, scaling to the X-ray surface flux for a solar-like star corresponding to $L_X=3\times 10^{30}$ yields $\dot{M}\sim 10^{-10}M_\odot$~yr$^{-1}$, only a factor of a few lower than our CME-based  estimate.  At a flux level of $\pi^1$~UMa, the relation corresponds to $\dot{M}\sim 10^{-11}M_\odot$~yr$^{-1}$---again similar to the CME scaling since the power law relations are also very similar.  

The scaling of the solar wind to higher magnetic activity levels remains a very uncertain endeavour owing to the lack of a comprehensive theory explaining the solar wind itself.   \citet{Cohen:11} argues that scaling mass loss according to X-ray luminosity is misleading because the latter is dictated by the closed magnetic field, while the former is dominated by open flux.  He argues that mass loss rates are unlikely to exceed $10^{-12}M_\odot$~yr$^{-1}$. The model of \citet{Cranmer.Saar:11} is driven by the energy flux of magnetohydrodynamic turbulence from the  convection zone and the filling factor  of open field.  Their mass loss rate for a saturated solar-like star is a few $10^{-12}M_\odot$~yr$^{-1}$, which is two orders of magnitude lower than the direct CME scaling.  
This suggests that the indirect wind observations of Wood et al.\  could in fact be observations of quasi-continuous CME mass loss for the more active stars of the sample, rather than a direct analogy to the solar wind.

Taken at face value, the solar CME data, combined with currently scant data on stellar winds and models of wind mass loss, suggest that at the highest activity levels mass loss could be dominated by CMEs, with a gradual transition to wind-dominated mass loss toward lower activities.   In the context of the study of \citet{Cohen:11}, unlike a steady wind, CME mass loss is expected to scale with X-ray luminosity because flares and CMEs generally originate from active regions that are dominated by closed field.  

In the context of the mass budget for individual CMEs, Equation~\ref{e:cme_mass} implies that for stellar flares with a total 1--8~\AA\ flare X-ray fluence of $10^{34}$~erg, the mean ejected mass is about $4\times10^{18}$~g.  If the CME source plasma resides in the corona, this mass is uncomfortably large.  It corresponds to the entire mass of a corona with a scale height of $0.1R_\odot$, an emission measure of a few $10^{52}$~cm$^{-3}$ (like that of 47~Cas~B from analysis of \citealt{Telleschi.etal:05}) and a quiescent active region-like density of $10^{10}$~cm$^{-3}$.   A fluence of $10^{34}$~erg, even limited to the 1--8~\AA\ GOES band, is still quite a modest flare compared with the largest flares seen on the most active stars and on T~Tauri stars, whose broad-band X-ray fluences can reach $10^{37}$~erg \citep[e.g.][]{Schrijver.etal:12}.  This corresponds to $\sim 3\times 10^{36}$~erg in the 1--8~\AA\ band based on the scaling derived in \S\ref{s:method}, and would imply a mean ejected mass a factor of 30 higher still.   These large mass requirements suggest that the solar CME X-ray fluence-mass relation must break toward the largest flares.

\subsection{Early Faint Sun Paradox}
\label{s:fys}

Could the mass loss through CMEs on an early active Sun be relevant to the 
``early faint Sun paradox''?   \citet{Sagan.Mullen:72}   pointed out that the lower solar luminosity predicted by stellar evolutionary theory earlier in the history of the solar system implies that for contemporary albedos and atmospheric composition global mean temperatures would have been below the freezing point of seawater until about 2.3~Gyr ago, in contradiction with geological evidence for liquid oceans.    Possible solutions to this paradox include higher concentrations of greenhouse gases and aerosols \citep[e.g.][]{Sagan.Mullen:72,Kasting:93}, a lower global albedo, either through less cloud coverage \citep[e.g.][]{Shaviv:03} or a smaller continental land mass \citep{Rosing.etal:10}.  An alternative solution is an early Sun more massive by several percent that has since been whittled down by mass loss \citep[e.g.][]{Guzik.etal:87,Sackmann.Boothroyd:03,Minton.Malhotra:07}.   

\citet{Wood.etal:02} note that their inferred steady wind mass loss rates are insufficient when combined with relations for the secular decline of X-ray surface flux:  the cumulative mass loss from an age of 1~Gyr or so is much less than 1\% .   As noted in \S\ref{s:mdot}, we find a relation between mass loss from CMEs and X-ray flux consistent with the observed steady wind relation of \citet{Wood.etal:02}---$\dot{M}\propto L_X^{1.5}$ for active stars, compared with their $\propto F_X^{1.34\pm {0.18}}$.   For a solar-like star such as  $\kappa^1$~Cet with an age of about 0.75~Gyr and $L_X\sim 10^{29}$~erg~s$^{-1}$, the mass loss rate is similar to that of $\pi^1$~UMa considered in \S\ref{s:mdot}, $\dot{M}\sim 3\times 10^{-12} M_\odot$~yr$^{-1}$.  Even if such a rate lasted for over a Gyr, it would amount to less than $0.01M_\odot$, or an order of magnitude less than required to resolve the  early faint Sun paradox unilaterally.   

\subsection{Energy loss, dynamo saturation and CME implications}
\label{s:ke}

A key question begging from the beginning of \S\ref{s:discuss} is what fraction of the stellar bolometric luminosity can be scavenged by magnetic processes that give rise to flares and CMEs? 
Saturation of magnetic activity for the most active stars pegs {\em broad-band} X-ray radiative losses at $\sim 10^{-3}L_{bol}$, which has generally been used as a saturation energy dissipation rate for convection zone dynamos.
However, X-rays represent only one aspect of the energy budget of solar flares.  The total radiative and non-thermal energy can be factors of 10-100 higher  than the GOES 1--8~\AA\ fluence, and possibly as much as associated CME kinetic energies \citep[e.g.][]{Woods.etal:04,Emslie.etal:05,Raymond:08,Kretzschmar.etal:10,Kretzschmar:11}.
This translates to factors of $\sim 3-30$ higher than broad-band X-ray fluence for flare-like temperatures.  If active stellar coronae are dominated by flares and their behavior is similar on active stars, then energy requirements of flares alone could amount to 1\%\ or more of the bolometric luminosity.   

We found in \S\ref{s:method} that scaling solar CME kinetic energy to a flare-dominated corona at  saturated activity level would require a staggering fraction of the stellar luminosity, approaching 10\%. 
If we declare 10\%\ of $L_{bol}$ too high a fraction of the total stellar energy budget to expend on CMEs, the implication is that the solar CME data {\em cannot} be extrapolated to significantly higher energies in the way we have done in \S\ref{s:method}.  Both CME speed and mass increase with X-ray fluence in the \citet{Yashiro.Gopalswamy:09} sample.  To avoid energy budget problems, either the CME kinetic energy vs.\ X-ray fluence must flatten out toward higher flare energies, implying that CME speed, mass, or both flatten out, or the CME-flare association rate must drop back significantly below 1.  It is commonly argued that flares without associated CMEs are confined by overlying magnetic field \citep[e.g.][]{Svestka.Cliver:92}.  
Active regions on active stars could confine more energetic CMEs associated with stronger flares because of stronger magnetic fields or different magnetic topology.

A fit to the mean CME speed, analogous to those in Eqns.~\ref{e:cme_mass} and \ref{e:cme_ke}, finds $v_c(E)=3.6\times 10^{-4}E^{0.22}$~km~s$^{-1}$, and for the same $10^{34}$~erg flare fluence corresponds to 11,000~km~s$^{-1}$.   The highest speed in the CME sample of \citet{Yashiro.Gopalswamy:09} is about 3000~km~s$^{-1}$.  The kinetic energy problem would be largely resolved were the ejection velocity to level out at this value toward higher flare energies.

The kinetic energy problem for very large CMEs is possibly related to the observed cutoff in solar energetic particle (SEP) fluence for particle energies above 10~MeV first inferred from cosmogenic radionuclides by \citet[][see also \citealt{Reedy:96,Hudson:07,Schrijver.etal:12,Usoskin.Kovaltsov:12}]{Lingenfelter.Hudson:80}.  The SEP fluence frequency spectrum breaks at approximately $10^{10}$~cm$^{-2}$.  Several different explanations have been suggested for this, such as an event energy dependence of SEP spectral distributions,  particle propagation effects in the heliosphere, or SEP opening angles depending on the energy of the triggering event  \citep[e.g.][]{Schrijver.etal:12}.  \citet{Hudson:07} notes that the SEP cutoff energy corresponds approximately to flares of X10 class (peak GOES 1-8~\AA\ flux of $10^{-3}$~Wm$^{-2}$ or $3\times 10^{27}$~erg~s$^{-1}$).  He argues that there appears to be no corresponding cutoff in the frequency of such flares, and any steeping of the flare frequency spectrum must happen at significantly higher energies.  This suggests there might be another limiting factor governing SEP production.
Since SEPs are thought to be largely accelerated in CME-driven coronal and interplanetary shocks \citep[e.g.,][]{Kahler.etal:78,Reames:99,Gopalswamy.etal:02,Cliver.etal:04}, a cutoff in energy could result from a corresponding limit to CME velocity or kinetic energy.   How such a constraint might translate to significantly more active stars would then be important for the CME kinetic energy budget: whether it scales with maximum flare, or active region, available energy, or is a more fundamental physical limitation with a cutoff at the same energy as seen on the Sun.  

We tentatively conclude that the relation between both CME mass and speed with flare fluence must flatten toward larger flare energies.  This behavior would appear to differ to the scaling of magnetic flux and flare properties from solar to active stellar cases.    \citet{Pevtsov.etal:03} find that, instead, the relationship between unsigned magnetic flux and X-ray spectral radiance for different regions of the Sun scales over 12 orders of magnitude to active stars.   Flare temperatures, emission measures and hard vs.\ soft components also appear to show a similar scaling from solar flares all the way to giant flares on active stars \citep{Feldman.etal:95,Isola.etal:07,Aschwanden.etal:08}.   Solar CME energies are limited by the free energy available in solar active regions, which \citet{Gopalswamy.etal:10} note is $< 10^{36}$~erg and usually at most $10^{33}$--$10^{34}$~erg \citep[e.g.][]{Emslie.etal:12,Aulanier.etal:13}; in this context it would be of interest to assess the energy available in stellar active regions, which might be possible through Zeeman-Doppler imaging magnetograms.

The large ``hidden" CME and flare energy requirements of a corona whose X-ray emission is dominated by flares implies that stars at the X-ray activity saturation threshold are extracting much more than $10^{-3}L_{bol}$ of energy, probably by an order of magnitude or more.   This suggests that saturated stars are experiencing a magnetic energy dissipation limit, rather than a limit imposed by the ability of the star to sustain X-ray emitting loops due to centrifugal stripping or poleward migration of magnetic flux \citep{Stepien.etal:01,Jardine.Unruh:99,Wright.etal:11}.  That is, 
the maximum amount of the total energy budget able to be extracted by a magnetic dynamo has been reached.  
If we allow CMEs to consume 1\%\ of the stellar energy budget on X-ray saturated stars, the implied mass loss for saturated stars is $\dot{M}_c\sim 5\times 10^{-11}M_\odot$~yr$^{-1}$.

\section{Conclusions}
Mean solar CME mass and kinetic energy are related to the associated flare X-ray fluence by power laws.  If active stellar coronal X-ray emission is comprised of flares as observations suggest, and these flares adhere to the observed solar flare-CME scalings found here, very high CME mass loss rates exceeding $10^{-10}M_\odot$ are implied for the most active stars, consuming a tenth of the stellar bolometric luminosity.   Since this energy requirement seems too high, we conclude that solar flare-CME relations cannot be extrapolated to arbitrarily high flare energies: CME mass and/or kinetic energy vs.\ flare X-ray fluence must flatten off at flare energies $\ga 10^{31}$~erg.  A more reasonable CME energy budget of 1\%\ of $L_{bol}$ implies $\dot{M}_c\sim 5\times 10^{-11}M_\odot$.  Even for budgets an order of magnitude lower it seems likely that mass loss from active stars will be dominated by CMEs.  The large ``hidden'' energy budget of flares and associated CMEs raises the question of what is the maximum amount of energy a solar-like star can extract from a magnetic dynamo?  If saturated stars are dominated by flares, this energy is likely to be an order of magnitude larger than the observed broad-band X-ray saturation level of $10^{-3}L_{bol}$.

\acknowledgments

JJD was funded by NASA contract NAS8-03060 to the {\it
Chandra X-ray Center} (CXC) and
thanks the CXC director, H.~Tananbaum, and the CXC science team
for continuing advice and support.  OC was supported by {\it Chandra} Grant TM2-13001X.  JJD also thanks David Soderblom for organising a workshop on the Faint Early Sun that provided the impetus for this study, and Vinay Kashyap for fruitful discussion.  Finally, we thank the referee for a very helpful report that enabled us to improve the manuscript significantly.


\clearpage

\end{document}